\documentclass[journal]{IEEEtran}

\usepackage{amsmath}
\usepackage{amssymb}
\usepackage{amsfonts}
\usepackage{hyperref}
\usepackage[plain]{fancyref}
\usepackage{graphicx, subcaption}
\usepackage{diagbox}
\usepackage{mathrsfs}
\usepackage{tabularx}
\usepackage{multirow}
\usepackage{tabulary}
\usepackage{booktabs}
\usepackage[version=3]{mhchem}
\usepackage{varwidth}
\usepackage{parnotes}
\renewcommand*\parnoteintercmd[1]{,\,}
\RequirePackage{lineno}
\usepackage{xcolor}
\usepackage{lipsum}
\usepackage{framed}
\usepackage{wrapfig}
\usepackage{amsmath}

\graphicspath{{./fig/}}

\begin{document}

\title{Mobile Object Tracking in Panoramic Video and LiDAR for Radiological Source-Object Attribution and Improved Source Detection}

\author{M.\,R.~Marshall, R.\,J.~Cooper, J.\,C.~Curtis, D.~Hellfeld, T.\,H.\,Y.~Joshi, M.~Salathe, K.~Vetter%

\thanks{This material is based on work supported by the US Department of Homeland Security under competitively awarded grant 16DNARI00026. Also, this material is based upon work supported by the Defense Threat Reduction Agency under DTRA 13081-31571. The development of the LEMURS system has been supported by the US Department of Homeland Security, Countering Weapons of Mass Destruction office, under competitively awarded contract/IAA 70RDND18K00000005. This support does not constitute an expressed or implied endorsement on the part of the United States Government.}%
\thanks{M.\,R.~Marshall and K.~Vetter are with the Nuclear Engineering Department at the University of California, Berkeley, Berkeley, CA 94720 USA.}%
\thanks{R.\,J.~Cooper, J.\,C.~Curtis, D.~Hellfeld, T.\,H.\,Y.~Joshi,  M.~Salathe, and K.~Vetter are with the Applied Nuclear Physics (ANP) Program at Lawrence Berkeley National Laboratory (LBNL), Berkeley, CA 94720 USA  (email: \href{mailto:thjoshi@lbl.gov}{thjoshi@lbl.gov}).}%
}

\maketitle

{---}

\begin{abstract}
The addition of contextual sensors to mobile radiation sensors provides valuable information about radiological source encounters that can assist in adjudication of alarms. 
This study explores how computer-vision based object detection and tracking analyses can be used to augment radiological data from a mobile detector system.
We study how contextual information (streaming video and LiDAR) can be used to associate dynamic pedestrians or vehicles with radiological alarms to enhance both situational awareness and detection sensitivity.
Possible source encounters were staged in a mock urban environment where participants included pedestrians and vehicles moving in the vicinity of an intersection.
Data was collected with a vehicle equipped with 6~NaI(Tl) $\bold{2~in.\times4~in.\times16~in.}$~detectors in a hexagonal arrangement and multiple cameras, LiDARs, and an IMU. 
Physics-based models that describe the expected count rates from tracked objects are used to correlate vehicle and/or pedestrian trajectories to measured count-rate data through the use of Poisson maximum likelihood estimation and to discern between source-carrying and non-source-carrying objects.
In this work, we demonstrate the capabilities of our source-object attribution approach as applied to a mobile detection system in the presence of moving sources to improve both detection sensitivity and situational awareness in a mock urban environment.
\end{abstract}
\section{Introduction}
\IEEEPARstart{R}{adiological} surveillance for gamma-ray emitting material in large-scale urban environments, such as city blocks, is an important mission in homeland security.
This involves searching for often weakly emitting gamma-ray sources in environments that are cluttered with pedestrians and vehicles, which makes the detection and localization of these sources extremely difficult.
In addition, when a source is detected and a radiological alarm occurs, alarm adjudication needs to be performed quickly due to the motion of the mobile detector system relative to objects in the scene. This can be challenging given the cluttered and dynamic nature of urban environments.
The addition of contextual sensors (e.g., streaming video and LiDAR) to a mobile detector system can provide valuable information about radiological source encounters that can assist in adjudication of alarms by providing associations between objects and the radiological data.
In this paper, we explore the concept of fusing contextual information from streaming video or LiDAR to augment radiation detectors on a mobile detector system. 

Previous methods have demonstrated the ability of simple contextual information (e.g. GPS and a camera) on a mobile detector system to improve situational awareness by overlaying a reconstructed 2D gamma-ray image onto a camera image \cite{wulf2008misti}.
More recent works have explored and demonstrated 3D gamma-ray imaging with free-moving
handheld devices~\cite{wahl2015polaris, haefner2017handheld, hellfeld2019real} by leveraging
advances in sensor and computational technology.
These methods utilize a set of contextual sensors, such as LiDAR and/or streaming video, in
conjunction with radiation sensors and algorithms that produce pose (position and orientation) estimates of
the free-moving device in a consistent reference frame.
All of the contextual sensor information, radiological data, and pose estimates are processed in real-time
to produce 3D visualizations of both the scene and gamma-ray image as the device moves through the environment.
While these methods can improve situational awareness by enabling 2D or 3D reconstructions of gamma-ray sources, they 
focus on static sources in stationary environments and currently are not well suited for dynamic environments with moving sources.
Thus, alarm adjudication by an operator in a cluttered environment with dynamic objects would
still be difficult to perform quickly and efficiently.

Recent approaches have used contextual-radiological data fusion to correlate trajectories from tracked objects with radiological data to better improve localization capabilities of moving sources compared to conventional reconstruction approaches.
Several works have used a LiDAR point cloud projected onto a XY-plane~\cite{henderson2020proximity, stadnikia2020data} in a constrained environment to correlate trajectories from 2D tracked objects with radiological data to attribute radiological sources to objects. 
Previous work by our group used advances in computer-vision-based object detection to detect physical objects (i.e., pedestrians and vehicles) using LiDAR and video independently in an environment with minimal constraints in extent~\cite{marshall2020three}. 
We then demonstrated the ability to reliably track the detected objects in 3D and to discern between source-carrying and non-source-carrying objects in a scene.
The findings from this work also suggested contextual information could be used to improve detection sensitivity. 
This work was performed using a static contextual sensor system with a co-located NaI(Tl) detector.
Here we build upon this previous work by applying this source-object attribution analysis concept to a mobile detector system equipped with video and LiDAR as well as six~$2~in.\times4~in.\times16~in.$~NaI(Tl) detectors in a hexagonal arrangement. 

Similar to the methods presented in our previous work~\cite{marshall2020three}, object detecting convolutional neural networks \cite{yolov4, yan2018} are applied to detect objects in image frames or LiDAR scans in real-time ($\sim$15~Hz), and a Kalman filter-based tracking algorithm with parameters specific to pedestrians and vehicles is used to track detected objects between video frames or LiDAR scans.
With our mobile system, it has been observed that the Kalman filter-based tracking algorithm performs more consistently if the detections are transformed into a static, consistent reference frame.
In this work, two different methods for computing pose information for the mobile system in a consistent reference frame are used, and we explore the impact these two methods have on tracking and attribution performance. 
Under the hypothesis that the object is associated with the source, we then generate models for a tracked object that describe the expected count rate from the tracked object in each detector within the detector array. 
Subsequently, the models for each detector are simultaneously fit to the observed count-rate data in each respective detector within the array. 
Finally, to identify the trajectories that are most (and least) likely to be associated with the radiological data, Poisson deviance is used as a goodness-of-fit metric~\cite{baker_clarification_1984}.

The source-object attribution analysis approach is independently evaluated using both video and LiDAR. 
Additionally, in our previous work, we introduced a track-informed integration window analysis to maximize the signal-to-noise ratio (SNR) for a tracked object.
This is extended here to a detector array. 
We identify time segments across the detector array that, when combined, will enable improved detection sensitivity compared to a summed response of the detector array (i.e., treating the detector array as a monolithic detector) or different fixed integration windows. 
We hereafter refer to identifying time segments across the detector array as an optimal configuration of detectors.

The article is structured as follows: the object detection, tracking, and source-object attribution analysis pipeline is discussed as well as the mobile system it operates on in Section~\ref{sec:methods}. 
In Section~\ref{sec:mobile_complex} and Section~\ref{sec:improve_det_sens_mobile_cases}, we demonstrate our source-object attribution analysis on a mobile system and improve detection sensitivity using tracking information in a mock urban environment with pedestrians and vehicles, respectively. 
Finally, a summary and future work are presented in Section~\ref{sec:conclusion}. \label{sec:intro}
\section{Methods}
\label{sec:methods}

For a more detailed discussion of the object detection, tracking, and analysis pipeline analysis approach refer to our previous work focusing on a static detection system~\cite{marshall2020three}. 

\subsection{LEMURS}
\label{sec:lemurs_}
\begin{figure}[htb!]
      \centering
        \includegraphics[width=3.2in]{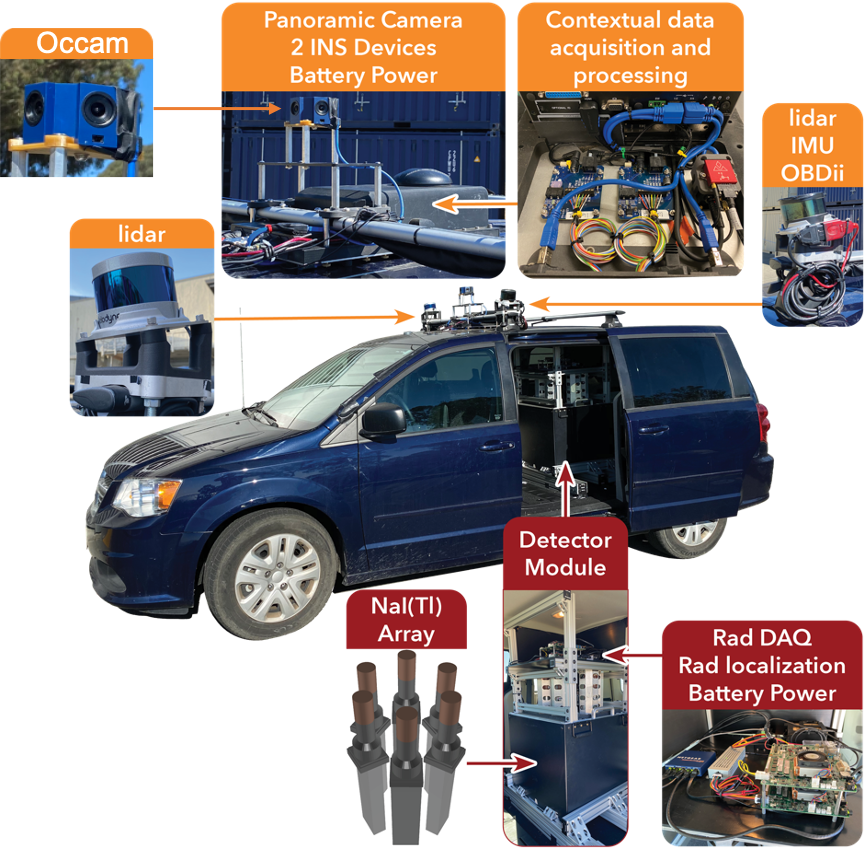}
\caption{The LEMURS system which consists of a panoramic Occam Omni~60 camera, 2 LiDAR, multiple IMU and INS devices, and six~$2~in.\times4~in.\times16~in.$~NaI(Tl) detectors.}
    \label{fig:lemurs_system}
\end{figure}

In this work, we used the LiDAR Enhanced Mobile Urban Radiation Search (LEMURS) vehicle~\cite{curtis_ieee} (\Fref{fig:lemurs_system}).
LEMURS consists of two 16-beam LiDAR mounted on the roof of the vehicle, multiple IMU and INS devices, a 360$^{\circ}$ panoramic Occam Omni 60 camera, and six~$2~in.\times4~in.\times16~in.$~NaI(Tl) detectors arranged in a hexagonal array.
Each detector is equipped with an Ortec DigiBASE \cite{digiBASE24} multichannel analyzer and configured to publish list-mode gamma-ray interaction data packets at 20~Hz with sub-millisecond synchronization between detectors.
The contextual sensor data and gamma-ray data are acquired within the Robot Operating System (ROS) \cite{ros} across multiple single-board computers with clocks synchronized by the network time protocol (NTP) \cite{ntp}.

\subsection{Object Detection and Tracking}
A lightweight, open-source object detector algorithm called you only look once (YOLO) was utilized to detect objects in the field of view (FOV) of the Occam camera \cite{yolov4} along with a ROS implementation, YOLO ROS \cite{yoloros}, which was modified to use a pre-trained YOLOv4-tiny object detection neural network \cite{yolov4}. 
From YOLO ROS, 2-D object detection bounding boxes are returned in the image coordinate system along with the object label and confidence score. 
Subsequently, the bounding boxes are converted to 3-D bounding boxes by inferring the distance of the object in the camera image. 
This is done by using both the camera intrinsic parameters and the object detection label.
For detected pedestrians and vehicles, nominal height information is assumed. 
A detected object label of person has an assumed nominal height of 1.75~meter~(m) \cite{fryar2016anthropometric}, and detected cars, trucks, motorcycles, and buses have nominal heights of 1.43~m, 1.80~m, 0.80~m, and 2.5~m, respectively. 

A separate object detection process is run for each Occam camera.
The camera frames are synchronized, and the object detections from all the cameras are collated.
This is done to avoid double-counting detections that take place in regions where the camera FOVs overlap.

To detect objects in LiDAR-generated point clouds, SECOND \cite{yan2018} with the PointPillars fast feature encoding \cite{lang2019} was used. 
Two 360$^{\circ}$ scans from each LiDAR are concatenated to produce sufficiently dense point clouds for inference using SECOND. 
To remove motion blur in the resulting point clouds, LEMURS LiDAR scans are transformed into a world-fixed frame before being aggregated.
The transformation to the world-fixed frame was found using two different methods.
In the first method, the pose estimates of the LEMURS vehicle in a world-fixed frame are calculated with Google Cartographer SLAM \cite{hess2016real} using the IMU and LiDAR data. 
The other method involved using a GPS stabilized by an INS \cite{spatial} to track the pose estimates of the LEMURS vehicle in a world-fixed frame. 
After aggregation, the point cloud is transformed back into the reference frame of LEMURS before inference because SECOND assumes the sensor is in the center of the frame. 
The timestamp used to map the point cloud back to the LEMURS reference frame is the average of all the timestamps from the point cloud ROS messages that were used to generate the aggregated point cloud.
The message format for the LiDAR detected object follows the same format as YOLO ROS. 
Additionally, it should be noted that SECOND provides 3-D bounding boxes because the depth of an object is directly measured with LiDAR and does not need to be inferred. 

Detected objects from both video and LiDAR are tracked using the modified Kalman-filter-based Simple Online Realtime Tracking (SORT) algorithm discussed in \cite{marshall2020three}. 
In the modified algorithm, 3-D bounding boxes are represented as multivariate normals (MVN) and the Hellinger Distance (HD) is used for data association, which is a measure of similarity between two MVNs and is scaled between 0 and 1 \cite{pardo2018statistical}. 
MVNs and HD are used for tracking over the traditional method of intersection-over-union (IOU) for two reasons. 
First, Kalman filters use MVNs to represent state variables, which we decided to leverage for tracking, and MVNs enable tracking across multiple cameras by allowing for uncertainty in object extent. 
The bounding boxes are represented as MVNs by making the center of each bounding box the mean of the MVN, and the off-diagonals of the MVN's covariance matrix are scaled dimensions of the bounding box dimensions.
Different scaled dimensions for LiDAR and video are used due to the fact that video has more uncertainty in an object's distance from the system since the information is inferred. 
Additionally, SECOND provides the heading (yaw) of bounding boxes, and this information was used to transform the off-diagonal elements of the covariance matrix.

Two state vectors -- position ($x$, $y$, and $z$) and velocity ($v_x$, $v_y$, and $v_z$) -- are tracked with our modified Kalman filter. 
By using the detected object's label, the velocity uncertainties in the covariance matrix are adapted according to the object's label to enable simultaneous tracking of both pedestrians and vehicles. 
We use 4.44 and 0.28 $m^2/s^2$ for vehicles and pedestrians, respectively. 
The velocity uncertainties and HD were found by running an optimization on scenes with either only vehicles or pedestrians present to find the lowest number of tracked objects produced while ensuring a limited number of incorrect associations by the data association algorithm. 

To perform data association between detections and their most likely corresponding track, the HD is calculated for each possible detection and track pair creating a matrix that has dimensions $D \times T$, where $D$ is the total number of detections
present and $T$ is the total number of available tracks. 
Linear assignment (also known as the Hungarian Method) is then applied to the matrix \cite{kuhn1955hungarian}.
If a detection and a track have a calculated HD of less than 0.8 they are consolidated to a single track. 
In addition, if two detections have a calculated HD of less than 0.6 when they are transformed into the world-fixed frame the two detections are consolidated to a single detection.
This is to prevent tracking the same object more than once.



With a mobile system, objects in the scene should be invariant to the motion of LEMURS which is not the case in a body-fixed frame.
To generate pose estimates of LEMURS in a world-fixed frame, INS or SLAM information is processed, and pose information is produced at a rate of 10~Hz.

It is thought that a navigational system that relies on GPS to produce pose estimates of the system's location in a global frame would have degraded tracking and attribution performance in an urban environment compared to applying SLAM.
Urban environments are cluttered with buildings that can occlude or reflect signals from GPS satellites to the LEMURS system.
To test this concept, the tracking and source-object attribution performance of using SLAM or a GPS stabilized by an INS (hereafter referred to as INS) to produce pose estimates of the LEMURS system in a world-fixed frame are compared.


\subsection{Modeling and fitting trajectories to Radiological Data}
When a radiological alarm occurs, the attribution analysis is triggered. 
In this case, a non-negative matrix factorization (NMF) based spectroscopic anomaly detection algorithm, described in \cite{bilton2019}, is used to determine the presence of an anomaly and perform source identification. 
The NMF-based anomaly detection algorithm is run independently on each detector within the detector array, and if a radiological alarm is triggered for any detector, the attribution analysis is triggered for all detectors. 
The attribution analysis is performed on all trajectories that are within 3~seconds of the start and stop of the radiological alarm.

The goal of the attribution analysis is to identify the trajectories that are most (and least) likely to have been associated with an alarm. 
This is done by assuming each track is responsible for the radiological alarm and modeling the expected count rate in each detector from a given track.
For a given discrete time step, $i$, the expected number of detected events, $c_i$ , within a spectral region of interest (ROI) centered at $E$, from a radioactive source with gamma-ray flux $\alpha$ in the presence of a constant background b can be described by

\begin{equation}
c_i(E) = \frac{\epsilon(\hat{\Omega}, E) \alpha e^{-\mu(E) \bold{r}_i}}{4 \pi \bold{r}_i^2} \cdot \Delta t_i + b \,,
\label{eq:rad}
\end{equation}

\noindent where $\epsilon$ is the effective area of the detector, $\bold{r}_i$ is the distance from the detector to the source, $\Delta t_i$ is a given integration time, and $\mu$ is an energy and medium dependent linear attenuation coefficient.
The effective area is a function of energy and the direction between the tracked object's position and the detector, $\hat{\Omega}$. 
The spectral ROI is defined using the isotope ID provided by the NMF alarming algorithm. 
The isotope ID, together with the direction between the object and the detector $\Omega$ at any given time is used to extract the appropriate $\epsilon$ from a look-up table of pre-computed response matrices.

A global best-fit model for each trajectory is found by simultaneously maximizing the Poisson likelihood between Eq.~\ref{eq:rad} for each detector and the observed count-rate data in each respective detector with a maximum likelihood estimation
algorithm \cite{Shepp1982}, where $\alpha$ and $b$ are free parameters. 

\begin{figure}[htb!]
   \centering
    \includegraphics[width=0.48\textwidth]{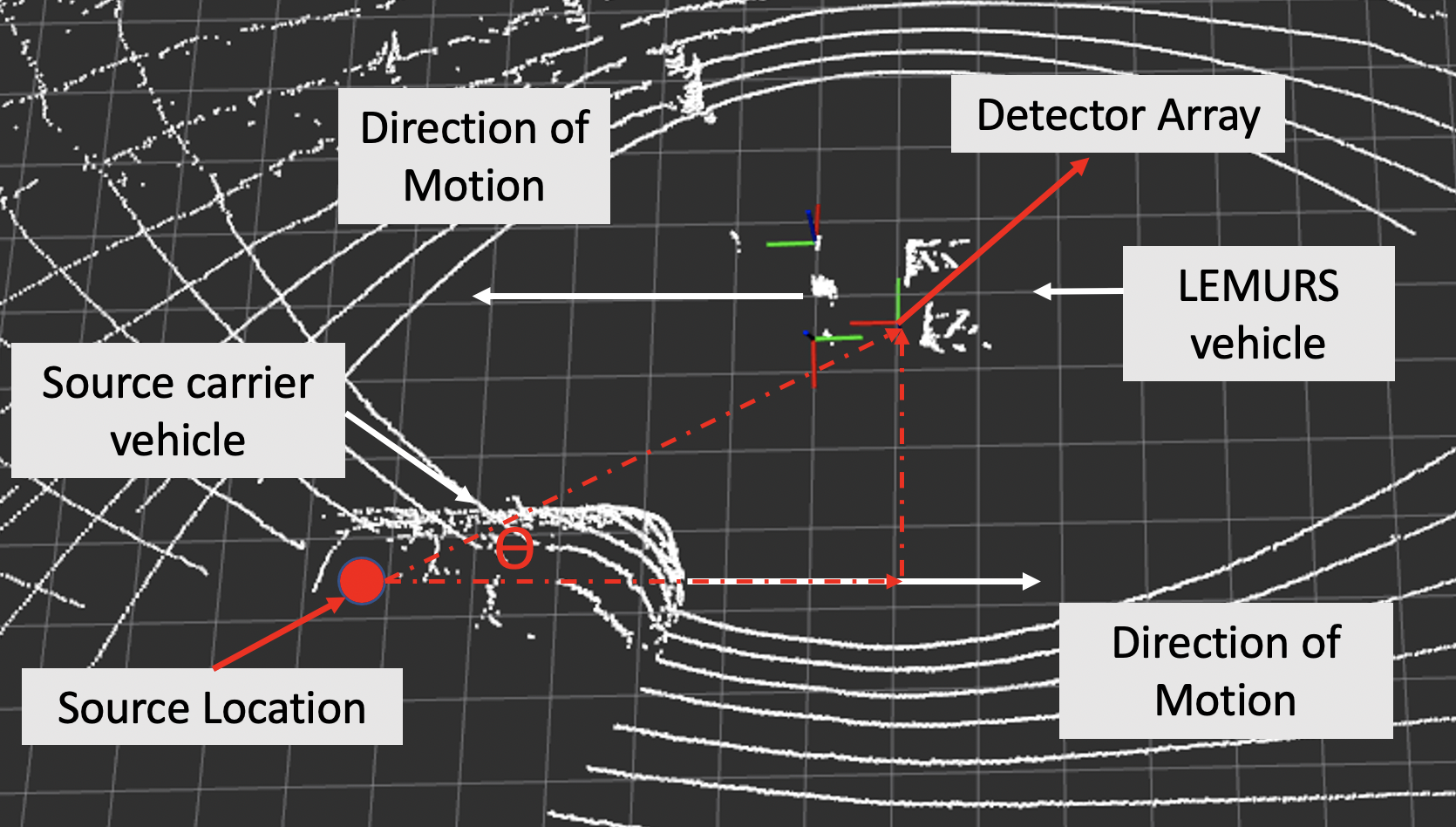}
    \caption{A top-down LiDAR image demonstrating how the estimated amount of attenuation present at a given angle $\Theta$ from a tracked object is determined.
    The tracking bounding boxes have been omitted for simplicity.
    The white dots and grid lines represent LiDAR returns and 1\,m$^2$ area, respectively, and the orientation of the two LiDARs and detector array on LEMURS are indicated by the red (x-axis), green (y-axis), and blue (z-axis) axes.}
    \label{fig:attenuation_present_}
\end{figure}

Details of accounting for the total attenuation coefficient, $\mu$, from tracked objects is described in more detail in~\cite{marshall2020three}. 
However, here we extend the calculation of $\mu$ to also include anisotropic attenuation from tracked objects. 
Depending on the location of a source within a tracked object, the amount of attenuation imposed by occluding material and shielding within the object can change throughout an alarm encounter as the source carrier and LEMURS drive past each other.
This has the potential to reduce the ability of the source-object attribution analysis to properly model the expected count rate, which could limit attribution performance as well as affect the track-informed integration window analysis described in Section~\ref{subsec:track_snr}.
To better handle anisotropic shielding from tracked objects, assumptions are made for both pedestrians and vehicles. 
With pedestrians, the radiological source is assumed to be in a backpack behind the object; whereas, with a vehicle, the source is assumed to be inside the vehicle. 
In order to model how attenuation changes as a function of angle relative to LEMURS, simulations were run using the Monte Carlo simulator MEGAlib~\cite{zoglauer2006megalib}. 
A pedestrian was modeled using the composition of a human~\cite{meyers2006dietary}, and a vehicle was modeled using a 1~m thickness of Aluminum (Al) for the engine block of the vehicle and 5~cm thickness of Al for the car doors and vehicle frame.
A source was placed either behind the pedestrian or centered in the vehicle's trunk. 
A NaI(Tl) detector was moved in 5$^\circ$ azimuthal increments around either object at a constant elevation in line with the source. 
The amount of attenuation present at each azimuth was calculated and applied to Eq.~\ref{eq:rad} for a given angle relative to LEMURS. 
\Fref{fig:attenuation_present_} displays an example of how the estimated amount of attenuation from a tracked object during an alarm encounter is determined.
It should be noted that the heading of each object is necessary for this calculation and thus can only be performed with LiDAR detected objects. 

\setlength{\tabcolsep}{0.5pt}

The energy- and direction-dependent response function was generated using MEGAlib~\cite{zoglauer2006megalib}.
The simulation included a detailed model of the detector array and a simplified model for the vehicle, operators, electronics, and mechanical supports~\cite{curtis}.
The gamma-ray response function was generated by modeling a radiological point-source located 10~m from the detector array center and 1.3~m off the ground, which corresponds to the elevation at the center of the detector crystals.
Separate simulations were performed with the source moved at 10$^{\circ}$ increments in azimuth around LEMURS for a total of 36 positions.
At each source position (i.e., direction), 2.7~$\times$~10$^{12}$ particles were simulated from an energy-dependent emission distribution normalized to source activity~\cite{mitchell2014gadras}.
For each detector, the effective area (per source activity) was computed using:

\begin{equation}
\epsilon = \frac{4 \pi \cdot R_{\mathrm{sim}}^2 \cdot X_{\mathrm{cnts}}}{N_{\mathrm{SimParticles}}},
\label{eq:eff_area}
\end{equation}

\noindent where $N_{\mathrm{SimParticles}}$ is the number of particles emitted into 4$\pi$, $R_{\mathrm{sim}}$ is the source distance from the detector, and $X_{\mathrm{cnts}}$ is the number of counts within the relevant peak-energy ROI.
Finally, to generate a 4$\pi$ response function, the response function was cosine modulated with elevation.
In Eq.~\ref{eq:eff_area}, it should be noted explicitly that $\epsilon=\epsilon(\hat{\Omega}, E)$.

\subsection{Attributing Trajectories to Radiological Data}
Similar to \cite{marshall2020three}, here we use the Poisson deviance to compute a p-value between the best-fit model for each trajectory and the count-rate data. 
Subsequently, a S-value ($\mathrm{log}_2(p)$) is calculated from the p-value and used to reject trajectories that are inconsistent with the data and are unlikely to be associated with the radiological alarm. 

The Kalman filter's localization prediction is based on the center of an object's bounding box so the source is assumed at the center of an object for the attribution calculation. 
To account for any position-source offsets that might exist, the modeled trajectory is calculated multiple times over a 0.5~second window while shifting the model trajectory by 0.1~seconds. 
The lowest S-value in this interval is used as the best-fit model for the trajectory. 
This approach does not account for potential offsets that exist in other dimensions (elevation and standoff). 

To reduce false positives associated with simply fitting background (i.e., incorrectly attributing the source to an object that is not responsible for the radiological alarm) in the attribution analysis, an additional metric is applied to reject trajectories. 
This is done by calculating Eq.~\ref{eq:rad} with and without the inclusion of $\alpha$ in the best-fit. 
Then, using the Bayesian Information Criterion (BIC) \cite{schwarz1978estimating}, it is determined whether a background only model ($\alpha$ set to 0 in Eq.~\ref{eq:rad}) or a source plus background model better describes the best-fit model. 
If a background only model better describes the best-fit model, the trajectory is most likely not responsible for the radiological alarm.
This method is described in more detail in \cite{marshall2020three}. 

\subsection{Improving Detection Sensitivity}
\label{subsec:track_snr} 
In our previous work~\cite{marshall2020three}, we demonstrated that detection sensitivity could be increased by using track-informed integration windows that optimize SNR. 
Here we expand upon this approach to include multiple detectors and within the detector array we identify an optimal configuration of detectors that will maximize SNR.
In addition, uncertainty in an object's extent is accounted for in the track-informed integration window formulation. 

The time-segments that, when combined, will maximize the expected SNR over a detector array for a given trajectory is found through the following analysis. 
Eq.~\ref{eq:max_SNR1} describes the expected SNR across $N$ discrete time windows for a tracked object within a fixed integration window $\Delta t$ at a point in time $i$ as

\begin{equation}
\mathrm{SNR} = \boldsymbol{\Bigg(}\displaystyle\sum_{i}^N s_i \Delta t_i\boldsymbol{\Bigg)}  \boldsymbol{\Bigg(}\displaystyle\sum_{i}^N b_i \Delta t_i\boldsymbol{\Bigg)}^{-1/2}\,
\label{eq:max_SNR1}
\end{equation}

\noindent where $s_i = \epsilon(\hat{\Omega}, E) \alpha e^{-\mu(E) r_i} / (4 \pi r_i^2)$ is
the photopeak count-rate, and $b_i$ is the mean background rate within $\Delta t_i$. 
Under the reasonable assumptions of constant source strength and background rate, $s_i/\sqrt{b_i}$ can be factored out of Eq.~\ref{eq:max_SNR1}. The sensitivity ($\$_T$) which is proportional to the $\mathrm{SNR}_T$ with the omission of $s_i/\sqrt{b_i}$ is found by 

\begin{equation}
\mathrm{SNR}_T \propto \$_T = \frac{\displaystyle\sum_{i \in T} \frac{\epsilon(\hat{\Omega}, E) \Delta t_i}{4\pi r_i^2} e^{-\mu(E) r_i}}{\sqrt{\displaystyle\sum_{i \in T} \Delta t_i}}, 
\label{eq:max_SNR3}
\end{equation}

\noindent where $T$ is a subset of time segments ($T\in[1,N]$) that when combined will maximize SNR for a given trajectory. 
In order to identify the optimal configuration of detectors within the array for a given alarm encounter, the models from each detector are concatenated together.
The subset of measurements $T$, that maximizes $\$$, is calculated using the concatenated data, and the spectra from each detector’s respective calculated optimal window are summed together to produce an optimal spectrum.

To account for the position uncertainties of each track in the optimal integration window formulation,
a Markov Chain Monte Carlo (MCMC) approach was applied~\cite{goodman2010ensemble} to the data to appropriately sample from the position uncertainties to better determine the optimal integration window.
MCMC is a method that draws samples directly from the posterior probability density function (PDF) distribution \cite{foreman2013emcee}.
MCMC does this by creating $M$ walkers that explore the parameter space and generate models of the data at each position.
The walker vector is defined by

\begin{equation}
\theta_i = \begin{pmatrix} x_i \\ y_i \\ z_i \end{pmatrix},
\label{eq:theta_vec}
\end{equation}

\noindent where $\theta_i$ is the estimated position at a discrete time step $i$ for a given trajectory.
The priors for each walker are the position uncertainty in $x_i$, $y_i$, and $z_i$ around each respective mean value.
The $\theta_i$ parameters are modeled to the count-rate data using Eq.~\ref{eq:rad}.
The source activity $\alpha$ and background are extracted from the best-fit model for the track, and the object position is varied to minimize the negative log-likelihood. 
The negative log-likelihood is determined by

\begin{equation}
\ell(\bold{x}|\boldsymbol{\lambda}) = [\boldsymbol{\lambda} -
\bold{x} \odot \log \boldsymbol{\lambda} + \log [\Gamma(\bold{x}+1)]]^{T} \cdot \bold{1},
\label{eq:negative_log_like}
\end{equation}

\noindent where $\odot$ denotes element-wise multiplication and $\Gamma(\cdot)$ is the gamma function. 
Also, $\boldsymbol{\lambda}$ is the best-fit model for a trajectory and
$\boldsymbol{x}$ is the count-rate over the span of the trajectory.

The initial guess for $\theta_i$ is the best-fit model for the track.
A total of 400 iterations with 600 walkers were run.
However, the first 100 iterations were
discarded because the walkers start close to the initial guess before fully exploring the parameter space.
This resulted in 180,000 samples, and a random subset of samples is chosen from the 180,000 samples to decrease the computational burden.
The total time duration of each optimal integration window is summed together across each detector for a given model.
The model that produces the largest time duration for the integration window across all six detectors but lowest negative log-likelihood within the subset is chosen.
It should be noted that once a model is chosen each detector’s respective optimal integration window is used, not the summed window, to identify the subset of time segments that when combined will maximize SNR.
Additionally, the number of detectors that contribute to the optimal configuration of detectors for a given source encounter can vary.
In the case of a single detector, the model that produces the largest time duration for the integration window but lowest
negative log-likelihood within the subset is chosen as the optimal window.

A spectroscopic analysis was applied to the track-informed optimal configuration of detectors and compared to either fixed integration windows (1.0, 2.0, 3.0, and 4.0~seconds) or a track-informed optimal integration window calculated from using the summed response of the detector array. 
For this analysis, the anomaly value -- the Poisson deviance between the observed data and a mean background spectrum scaled to match the observed counts -- computed by the NMF-based anomaly detection algorithm was used as a proxy for detection sensitivity where a larger anomaly value suggests improved detection sensitivity through this track-informed analysis. 
\section{Source-object Attribution in a Mock Urban Environment}
\label{sec:mobile_complex}
\begin{figure}[htb!]
   \centering
    \includegraphics[width=0.48\textwidth]{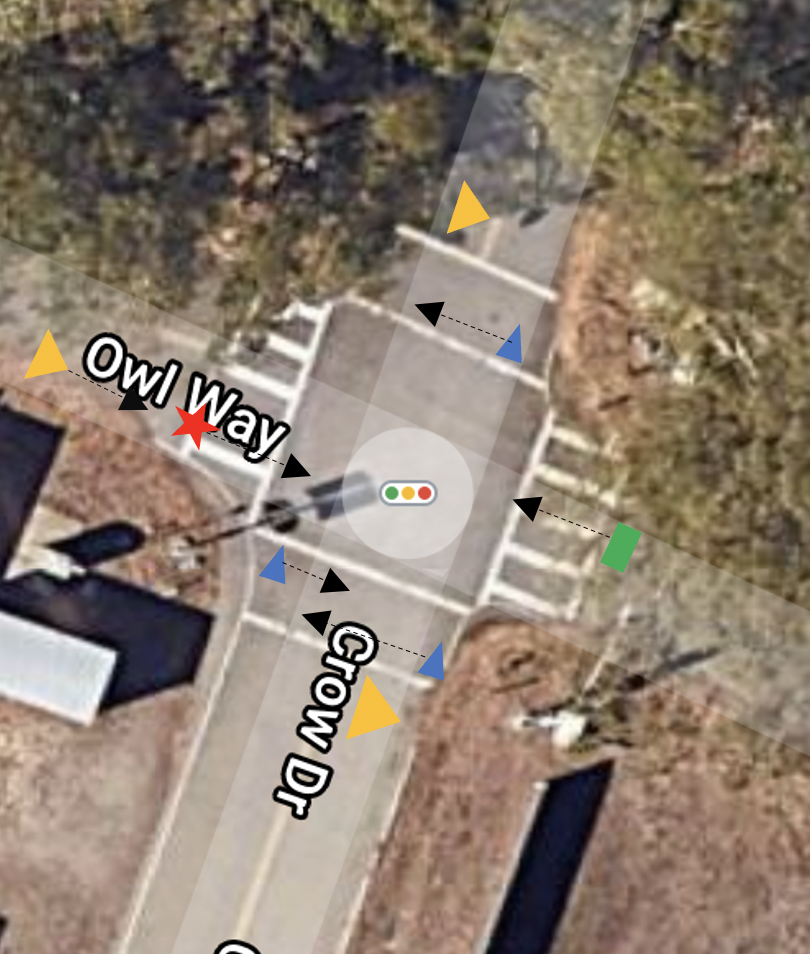}
    \caption{A top-down view of the mock urban environment intersection along with the objects present and their direction of motion. 
    The orange right-facing triangles, blue triangles, red star, or green square indicate vehicles, pedestrians, the source carrier, or LEMURS, respectively.
    The arrows indicate the direction of motion for each object during the alarm encounters. 
    If an arrow is not associated with an object, the object was stationary.}
    \label{fig:depiction_of_scene}
\end{figure}

The performance of the source-object attribution analysis in the presence of a mobile source was tested at a mock urban environment at Richmond Field Station (RFS).
A 1.87~mCi $^{137}$Cs source inside of 2~cm of lead-shielding was placed in the trunk of a vehicle and driven around. 
LEMURS and the source carrier performed straight line drive-bys going either 10 or 20~mph past each other.
\Fref{fig:depiction_of_scene} depicts the location of the objects present and their direction of motion overlaid on a top-down view of the intersection~\cite{google}. 
Both LEMURS and the source carrier drove straight for 15~m before passing in the middle of an intersection that had pedestrians on either side of the intersection walking parallel to LEMURS and the source carrier.
Two stationary cars were on both sides of the intersection and were perpendicular to the direction of motion of LEMURS.
Additionally, the source carrier was followed by a car traveling 10 or 20~mph depending on the scenario.
These scenarios were repeated at least 18 times for both speeds, and the lowest exclusion metric for a given alarm encounter is used as a metric to determine attribution performance.
Attribution was performed using the photopeak ROI (600~keV -- 725~keV) for $^{137}$Cs.

\setlength{\tabcolsep}{0.5pt}

\subsection{10~mph Scenario}
\label{sec:10mph_scenario}
\begin{figure}[htb!]
   \centering
   \begin{subfigure}[b]{2.6in}
         \includegraphics[width=\textwidth]{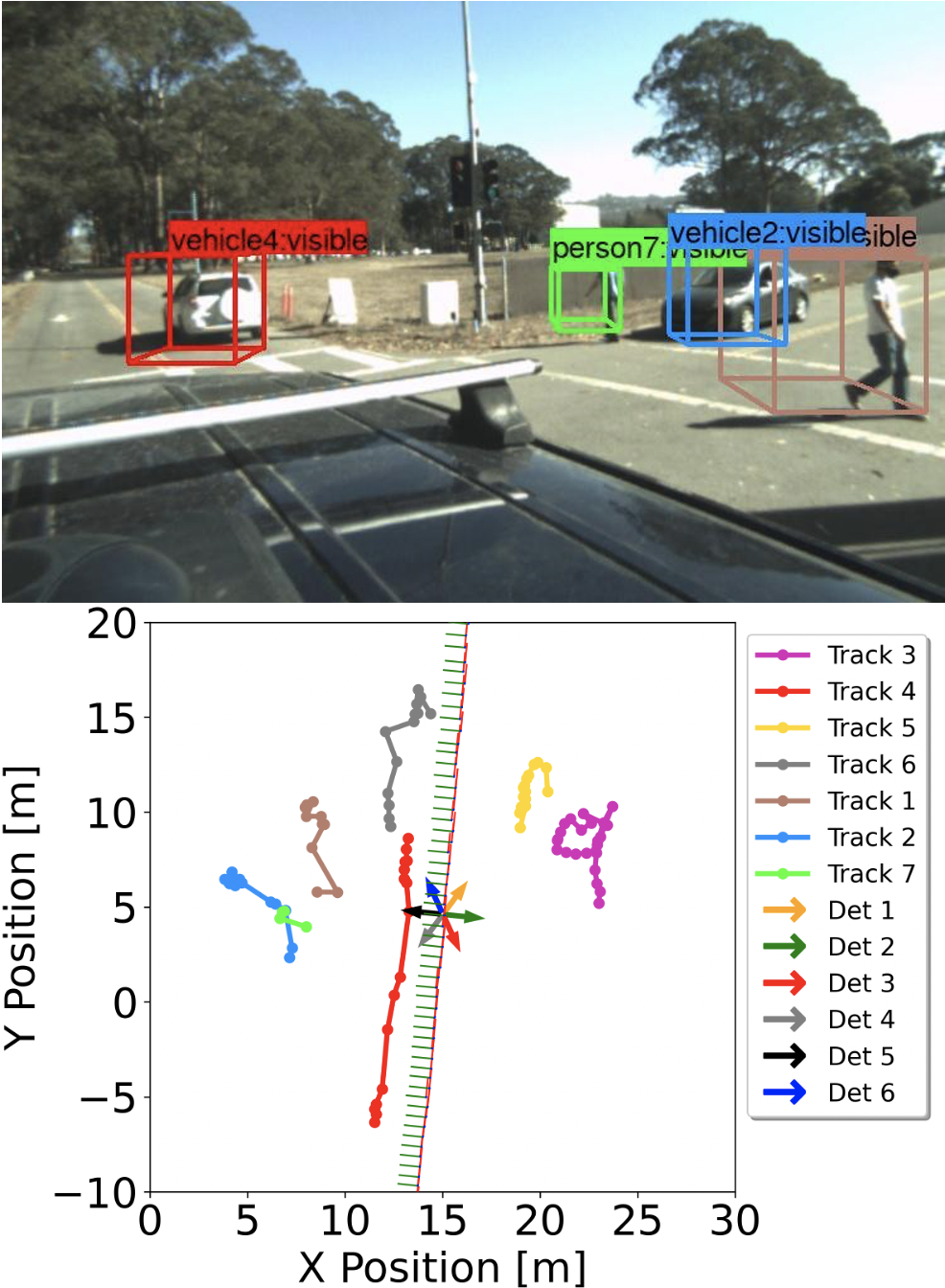}
           \caption{}\label{fig:10mph_complex_tracker_video}
           \end{subfigure}
       \begin{subfigure}[b]{2.8in}
       \centering
         \includegraphics[width=\textwidth]{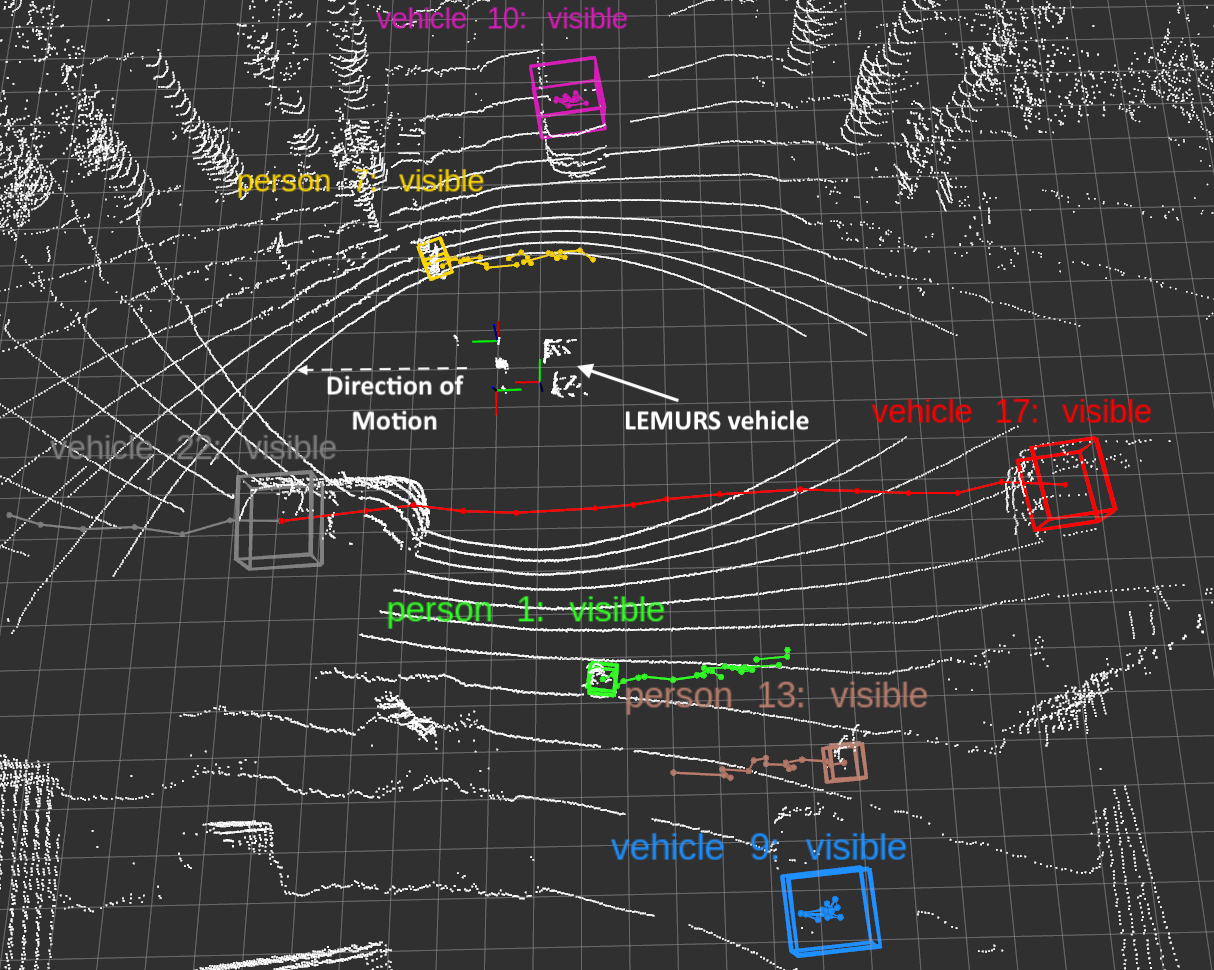}
         \caption{}\label{fig:10mph_complex_tracker_lidar}
     \end{subfigure}

    \caption{Example of object tracking in a mock urban environment using video (a) and LiDAR (b) data with the LEMURS system driving 10~mph past a vehicle carrying a 1.87~mCi~$^{137}$Cs source inside of 2~cm Pb shielding and using SLAM to create pose estimates of LEMURS in a world-fixed frame. 
    (a) shows the output of the tracker analysis for one camera image (top) and the trajectories for each object along with the orientation of LEMURS indicated by the red (x-axis), green (y-axis) (bottom).
    Also, in the bottom pane, the detector orientations are shown.
    In (b), the same alarm encounter as (a) using LiDAR point clouds is shown.
    The bounding box colors of the objects (but not the labels) are consistent with (a).
    In (b), the trajectory of each object to that point is shown.
    The white grid lines represent 1\,m$^2$, and the orientation of the two LiDARs and detector array on LEMURS are indicated by the red (x-axis), green (y-axis), and blue (z-axis) axes.}
    \label{fig:10mph_complex_tracker}
\end{figure}

\Fref{fig:10mph_complex_tracker} shows the output of the tracking analysis from one alarm encounter using video (\Fref{fig:10mph_complex_tracker}a) and LiDAR-based trajectories (\Fref{fig:10mph_complex_tracker}b) for a LEMURS and source carrier speed of 10~mph and SLAM to create pose estimates of LEMURS in a world-fixed frame. 
In \Fref{fig:10mph_complex_tracker}a (top pane), the progression of the alarm encounter as the source carrier (Track~4 - white vehicle) drives past LEMURS using video data along with the trajectory for each object (bottom pane) is shown. 
\Fref{fig:10mph_complex_tracker}b shows the same alarm encounter and moment in time as the image in the top pane of \Fref{fig:10mph_complex_tracker}a but using LiDAR trajectories.
The color-coding of each object (but not the labels) has been kept consistent with \Fref{fig:10mph_complex_tracker}a. 
For both video and LiDAR, the source carrier and surrounding objects are continuously tracked throughout the alarm encounter. 
One can see the inherent depth information extracted from LiDAR enables more reliable position estimation compared to the inferred distance estimation for video, which adds noise in the trajectories. 
In the bottom pane of \Fref{fig:10mph_complex_tracker}a, both Track~6 and Track~2 were stationary vehicles throughout the encounter, but due to frame-to-frame uncertainty in the distance estimate, the Kalman filter pose estimation for both objects varies; whereas, the variability of Track~11 and Track~13 (Track~2 and Track~3 from \Fref{fig:10mph_complex_tracker}a bottom pane) in \Fref{fig:10mph_complex_tracker}b is more concentrated around each object's respective position.

Additionally, in \Fref{fig:10mph_complex_tracker}a (bottom pane), the detector orientation of LEMURS is seen. 
In this alarm encounter, Detectors~4-6 are the detectors closest to the source carrier during the time of closest approach.

\setlength{\tabcolsep}{0.5pt}

\begin{figure}[htb!]
   \centering
       \begin{subfigure}[b]{3.4in}
       \centering
         \includegraphics[width=\textwidth]{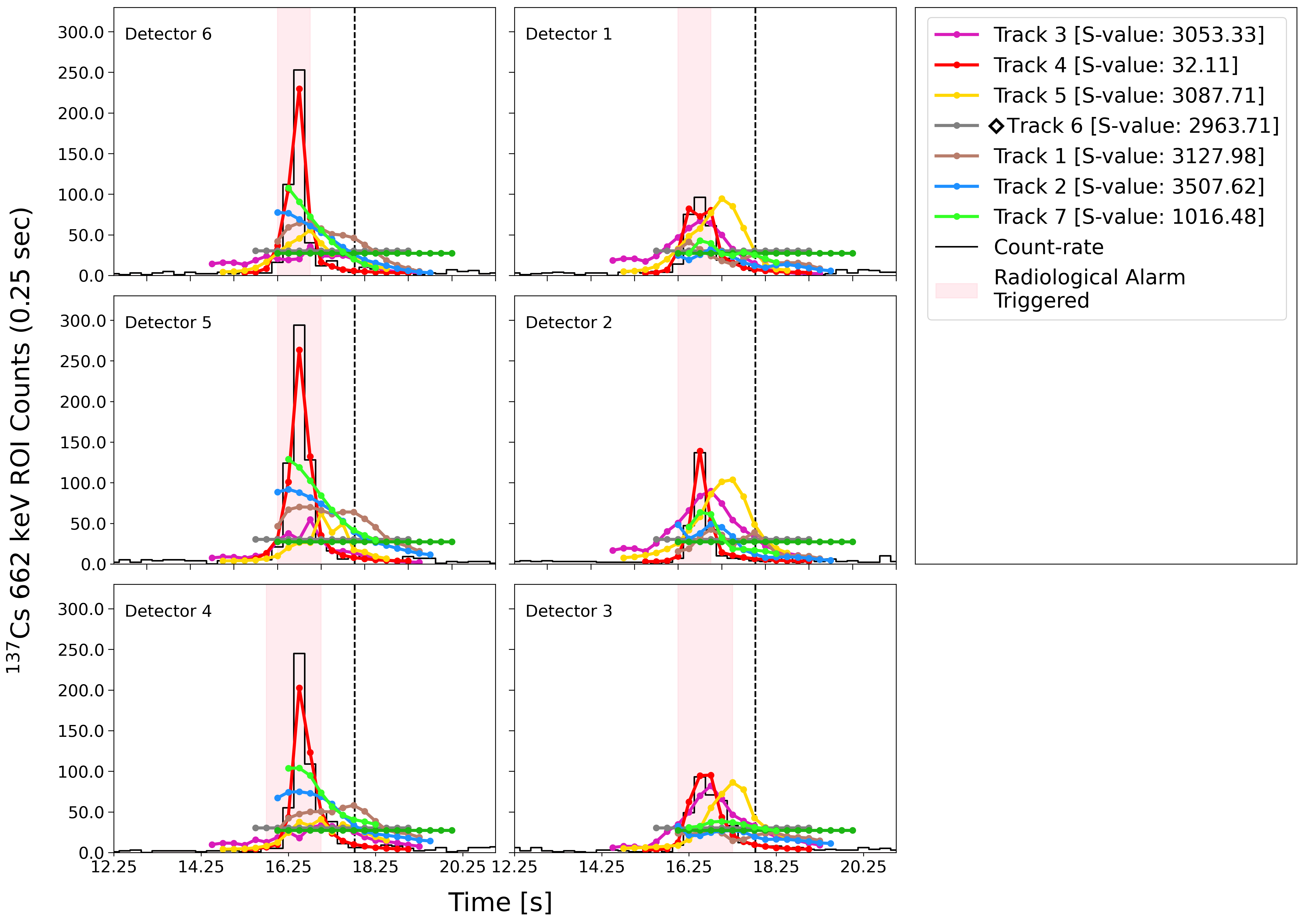}
         \caption{}\label{fig:complex_video_10mph_c}
     \end{subfigure}
     \quad
    \begin{subfigure}[b]{3.4in}
      \centering
        \includegraphics[width=\textwidth]{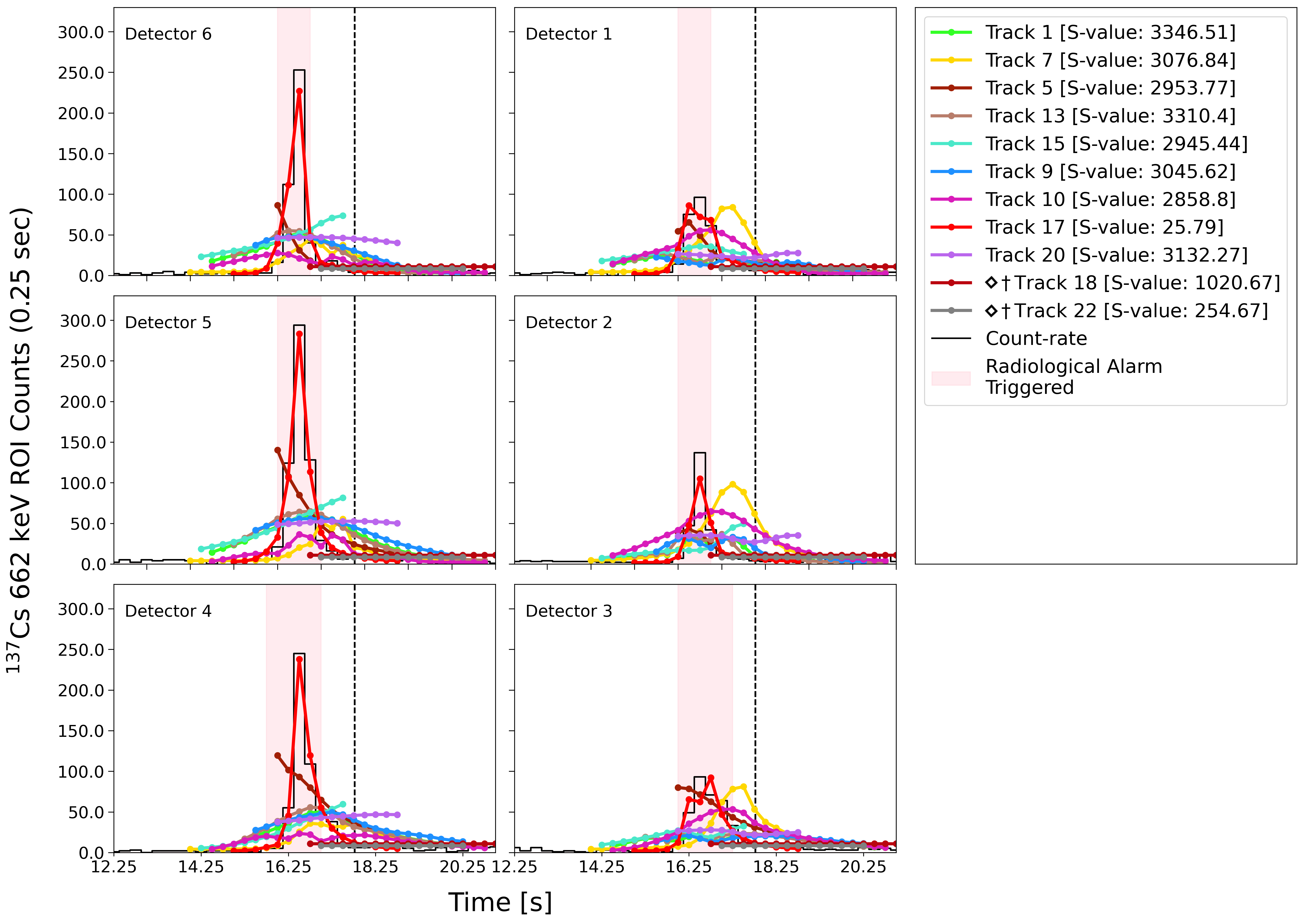}
        \caption{}\label{fig:complex_LiDAR_10mph_c}
    \end{subfigure}

    \caption{The result of the alarm encounter depicted in \Fref{fig:10mph_complex_tracker} using source-object attribution with video (a) and LiDAR-based trajectories (b) in a mock urban environment.
    In (a-b), the result of the alarm encounter for video and LiDAR are shown, respectively.
    The arrangement of the count-rate data from each detector matches the arrangement of the detectors in the LEMURS array.
    The diamond indicates the best-fit model was better described by a background only model, and the dagger indicates more than 95\% of the trajectory was outside of the radiological alarm.
    Also, the dashed line in (a) and (b) correspond with the moment in time depicted in \Fref{fig:10mph_complex_tracker}a (top pane) and \Fref{fig:10mph_complex_tracker}b, respectively.}
    \label{fig:10mph_complex_single_encounter}
\end{figure}

The results of the alarm encounter using video and LiDAR-based trajectories are shown in \Fref{fig:10mph_complex_single_encounter}a and \Fref{fig:10mph_complex_single_encounter}b, respectively. 
The orientation of the detector panes in both figures matches the detector orientation of LEMURS shown in the bottom pane of \Fref{fig:10mph_complex_tracker}.
In \Fref{fig:10mph_complex_single_encounter}a, the 1/r$^2$ profile in the count rate is seen from LEMURS and the source carrier driving past each other along with the different best-fit models for all the objects present during the alarm encounter. 
The best-fit model for Track~4 (white vehicle) clearly follows the radiological data, and this is the correct attribution as the white vehicle was responsible for the radiological alarm.
Track~5 has some correlation with Detectors~1-3, but with the angular response of the detector array, the trajectory does not follow the count-rate data in Detectors~4-6 and can be excluded.
The remaining trajectories can also all be excluded from the analysis. 
A similar result is seen using LiDAR data.  
All of the trajectories can be excluded from the analysis except for Track~17, which correlates with the radiological data and is the correct attribution. Additionally, for video and LiDAR-based trajectories, the calculated time offset between the estimated Kalman filter position and source location was 0.2~seconds (1~meter) or 0.25~seconds (1.3~m), respectively, which corresponds with the trunk of the source carrier's vehicle. 
Thus, both the source-carrying object and the source location within the object were correctly identified using our source-object attribution analysis with either video or LiDAR data.  

\setlength{\tabcolsep}{0.5pt}

With LiDAR, there are more trajectories than objects present in the scene.
The LiDAR detection CNN has a higher number of false positive detections (i.e., detecting an object when in fact no object is present at the given location) compared to Yolov4-tiny.
These artifacts could be limited by increasing the minimum confidence score needed to track an object.
However, increasing the minimum confidence score could decrease tracking performance for certain objects.
The false positive detections would need to be discarded by an operator in real time.
Additionally, sensor fusion, such as combining LiDAR and video, could be used to discard spurious detections, but sensor fusion is outside the scope of this paper.

\begin{figure*}
     \centering
     \includegraphics[width=\textwidth]{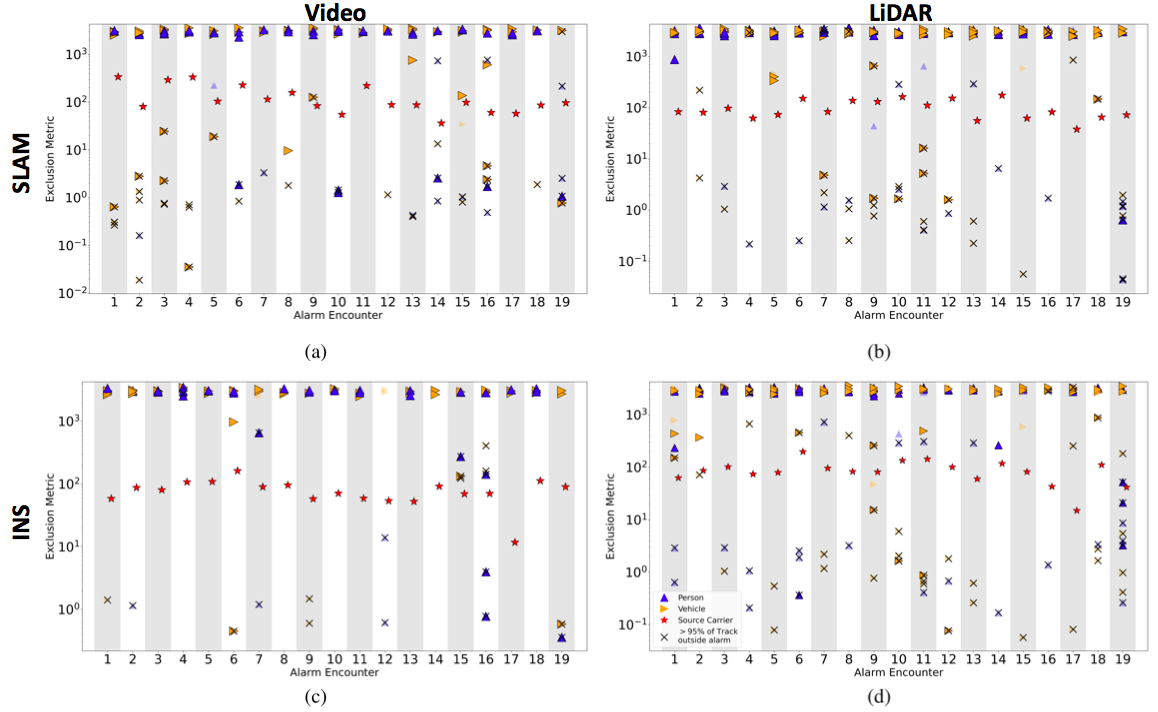}
    \caption{Source-object attribution for all alarm encounters when LEMURS drove 10~mph past a vehicle traveling 10~mph and carrying carrying a 1.87~mCi~$^{137}$Cs source inside of 2~cm Pb shielding in a mock urban environment using video (a, c) and LiDAR (b, d). In both (a) and (b), SLAM was used to generate a consistent reference frame, and in (c) and (d), INS data was used. 
    In (a) -- (d), the faded (enlarged, outlined) points indicate best-fit models that were better described by a background only (source plus background) model.
    The black x's indicate tracks that are 95\% or more outside of the
    radiological alarm.}
    \label{fig:10mph_complex}
\end{figure*}

In \Fref{fig:10mph_complex}, all of the alarm encounters at 10~mph are shown using SLAM for tracking. 
For a given alarm encounter, the exclusion metric for each object present along with the object's label is displayed. 
For example in \Fref{fig:10mph_complex}a, Alarm Encounter~2 has 10 objects present during the encounter. 
Five of the objects have lower exclusion metrics than the source carrier; however, these objects either have best-fit models where BIC preferred a background only model or more than 95\% of the track is outside of the radiological alarm so these tracks can be rejected from the analysis.
Thus, the source carrier has the lowest exclusion metric among relevant best-fit models.

In all of the trials, the source is correctly attributed to the source carrier in 18 out of 19 trials (\Fref{fig:10mph_complex}a) using video, and for LiDAR-based trajectories (\Fref{fig:10mph_complex}b), the source carrier is correctly identified in 19 out of 19 trials.
The effective attribution for both video and LiDAR throughout all the alarm encounters is enabled by effective tracking of objects.
In the one trial where the source carrier did not have the lowest exclusion metric for video, an obvious correlation between the best-fit model and radiological data existed.
An operator monitoring in real time would be able
to correctly attribute the radiological alarm to the source carrier. 
Additionally, for both video and LiDAR, the average time offset applied to the estimated source position was 0.35~seconds (1.5~m) locating the source to the rear of the source carrier's vehicle. 
These results demonstrates how the source-object attribution analysis can bring situational awareness to a mobile detector system.

The results of tracking using pose estimates produced with the INS for all the alarm encounters for video and LiDAR are shown in \Fref{fig:10mph_complex}c and \Fref{fig:10mph_complex}d, respectively.
The results are similar to tracking and performing source-object attribution using SLAM.
In this case, the source carrier has the lowest exclusion metric in 19 out of 19 trials using both LiDAR and video-based trajectories.
Also, similar to the SLAM results, the source was localized to the trunk using both video and LiDAR. 

The results from this analysis do not match the hypothesis, that INS should produce degraded pose estimates in an urban environment, which would adversely impact the source-object attribution analysis.
In a typical urban environment, buildings will reflect and block satellite signals causing signal interference reducing the position accuracy in the pose estimate.
With navigational systems, this loss of position accuracy from obstructions is expressed as dilution of precision (DOP).
The fewer satellites available for the pose estimate, the higher the DOP value.
In the mock urban environment considered here the vertical DOP (VDOP) and horizontal DOP (HDOP) values range from 1--20, respectively, where values ranging from 5--20 indicate moderate to low confidence levels in the pose estimates due to high environmental interference.
The larger DOP values are due to a high number of tall trees in the mock environment that occlude the satellite signal.
While the INS does filter between GPS coordinates at $\sim$1~Hz to improve GPS accuracy and reduce jitter in the pose estimates, in these environments the position accuracy is still reduced.
When the INS pose information for the alarm encounters is overlaid onto a map of the area there is an obvious drift over time of the pose estimates (i.e., the pose estimates for LEMURS do not correspond with the road).
However, for a given alarm encounter, all the objects in the scene are relative to this drift since the objects are transformed into the world-fixed frame.
Thus, in the 10~mph scenario, tracking and attribution can still be performed effectively and INS performance is similar to using SLAM, but we expect performance with an INS to worsen in an environment subject to degraded GPS performance (e.g., an urban canyon).

Overall, using our source-object attribution analysis in these alarm encounters, an apparent connection between the radioactive source origin and the detected signal existed, and we were able to both correctly localize the source to the object responsible for the radiological alarm and correctly localize the position of the source within the object. 
This was demonstrated using both video or LiDAR data and either INS or SLAM to generate pose estimates of LEMURS in a world-fixed frame. 
In all of these alarm encounters, an operator would be able to quickly and effectively perform alarm adjudication.

\begin{figure*}[htb!]
     \centering
      \includegraphics[width=\textwidth]{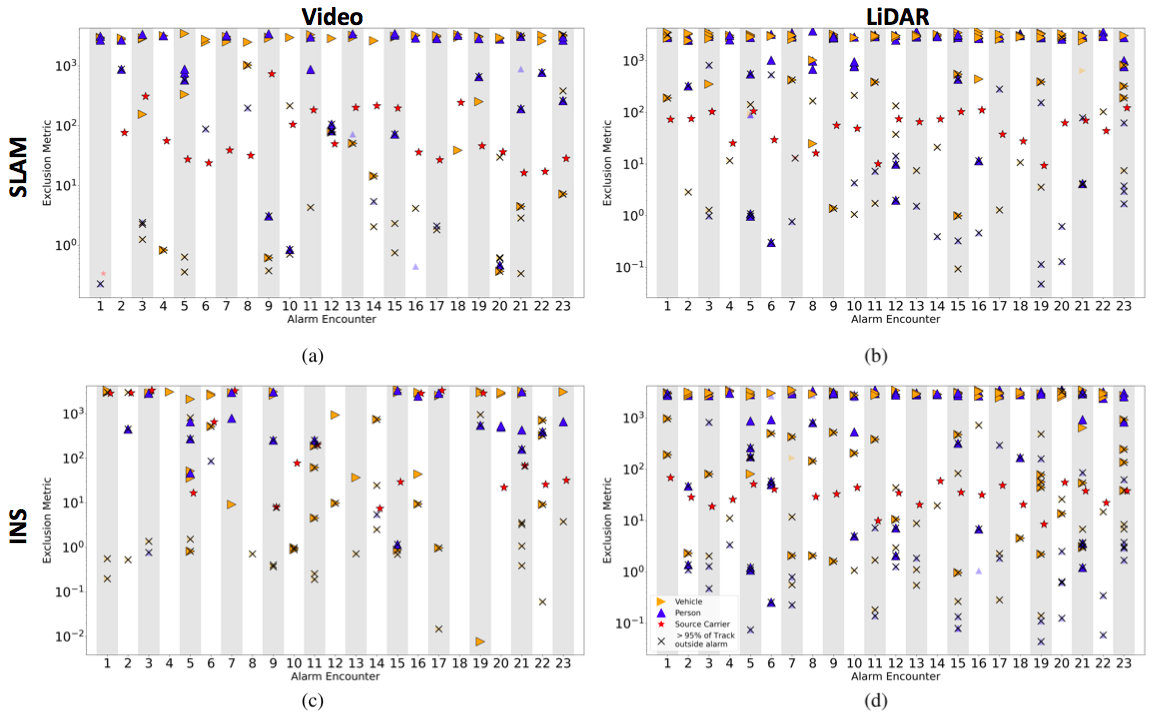}

    \caption{Results from performing source-object attribution on all radiological alarm encounters when both LEMURS and the source carrier were traveling 20~mph relative to each other in the mock urban environment using video (a, c) and LiDAR (b, d). 
    SLAM [INS] was used in (a) and (b) [(c) and (d)] to generate pose estimates in a world-fixed frame.
    In both (a) -- (d), the faded (enlarged, outlined) points
    indicate best-fit models that were better described by a background only (source plus background) model.
    The green circles indicate tracks that are 95\% or more outside of the radiological alarm.}
    \label{fig:20mph_complex}
\end{figure*}

\subsection{20~mph Scenario}
\label{sec:20mph_scenario}
The same scenario presented in Section~\ref{sec:10mph_scenario} was repeated for a LEMURS and source carrier speed of 20~mph.
These trials were performed to better understand tracking and attribution performance at higher vehicle speeds that are more relevant for urban environments.
In this case the scenario was repeated 23 times.

The results from all the alarm encounters for the 20~mph scenario using SLAM are shown in \Fref{fig:20mph_complex}a and \Fref{fig:20mph_complex}b for
both video and LiDAR, respectively.
Using video (LiDAR), the source carrier was correctly attributed to the radiological data in 20 (23) out of 23 trials while the remaining trajectories could be rejected. 
For both video and LiDAR, the average source offset was 0.15~seconds (1.3~m), which correctly located the source in the rear of the source carrier's vehicle.
Using video, in one of the trials, the source carrier was inconsistently detected throughout the alarm encounter and no attribution between the object responsible for the radiological alarm and the count-rate data was made.
For the remaining two trials with video, inconsistent tracking of non-source-carrying objects during the period of time in which the radiological alarm was triggered resulted in lower best-fit models for those objects than the source carrier; nonetheless, a clear correlation existed between the count-rate data and the best-fit model for the source carrier and an operator monitoring in real time would be able to identify the object responsible for the radiological data. 
The results from these encounters demonstrate the flexibility of a Kalman filter and the advantage of using MVN tracking.
In these alarm encounters, objects were only detected and tracked for a short period of time; however, for both video and LiDAR, all the objects were continuously tracked throughout a majority of the transient alarm encounter, which enabled more effective alarm attribution in these cases.

Tracking using only INS data, the source carrier has the lowest exclusion metric in 10 out of 23 alarm encounters with video-based trajectories (\Fref{fig:20mph_complex}c) and 22 out of 23 alarm encounters with LiDAR-based trajectories (\Fref{fig:20mph_complex}d).
With video, there were 13 alarm encounter where alarm attribution could not performed due to inconsistent tracking.
This result is in contrast with the video SLAM results from \Fref{fig:20mph_complex}a where the source carrier had the lowest exclusion metric in 20 out of 23 alarm encounters.
Since the uncertainty in a detected object's distance is consistent using either INS or SLAM with video data, the discrepancy between INS and SLAM results in the 20~mph scenario appears to be driven by increased noise in the INS pose information at greater speeds.
This increased noise is sufficient enough along with the inferred distance information to cause degraded tracking and attribution performance for video.
The LiDAR results using INS are comparable to the SLAM results due to more consistent depth information. 

\section{Improved Detection Sensitivity with Track-informed Optimized Integration Windows}
\label{sec:improve_det_sens_mobile_cases}
With tracking information, optimal integration times to maximize SNR can be found using the formulation discussed in Section~\ref{subsec:track_snr}.
Furthermore, for a given alarm encounter, certain detectors will be closer to the source carrier and will experience higher SNR than the detectors further from the source carrier.
Thus, an optimal configuration of detectors should exist that maximizes SNR.
The following two sections (Section~\ref{sec:10mph_opt_win} and
Section~\ref{sec:20mph_opt_win}) investigate optimizing integration windows to maximize SNR using the experimental alarm encounters discussed in Section~\ref{sec:10mph_scenario} and Section~\ref{sec:20mph_scenario}.
Optimal integration windows are found either using the 6~NaI detectors independently or summing the response of the 6~NaI detectors.
In both cases, position uncertainty of the object is accounted for with MCMC.
Additionally, the analyses were only performed using the trajectories generated from tracking with SLAM.

\subsection{10 mph Scenario}
\label{sec:10mph_opt_win}

\begin{figure*}[htb!]
     \centering
      \includegraphics[width=0.9\textwidth]{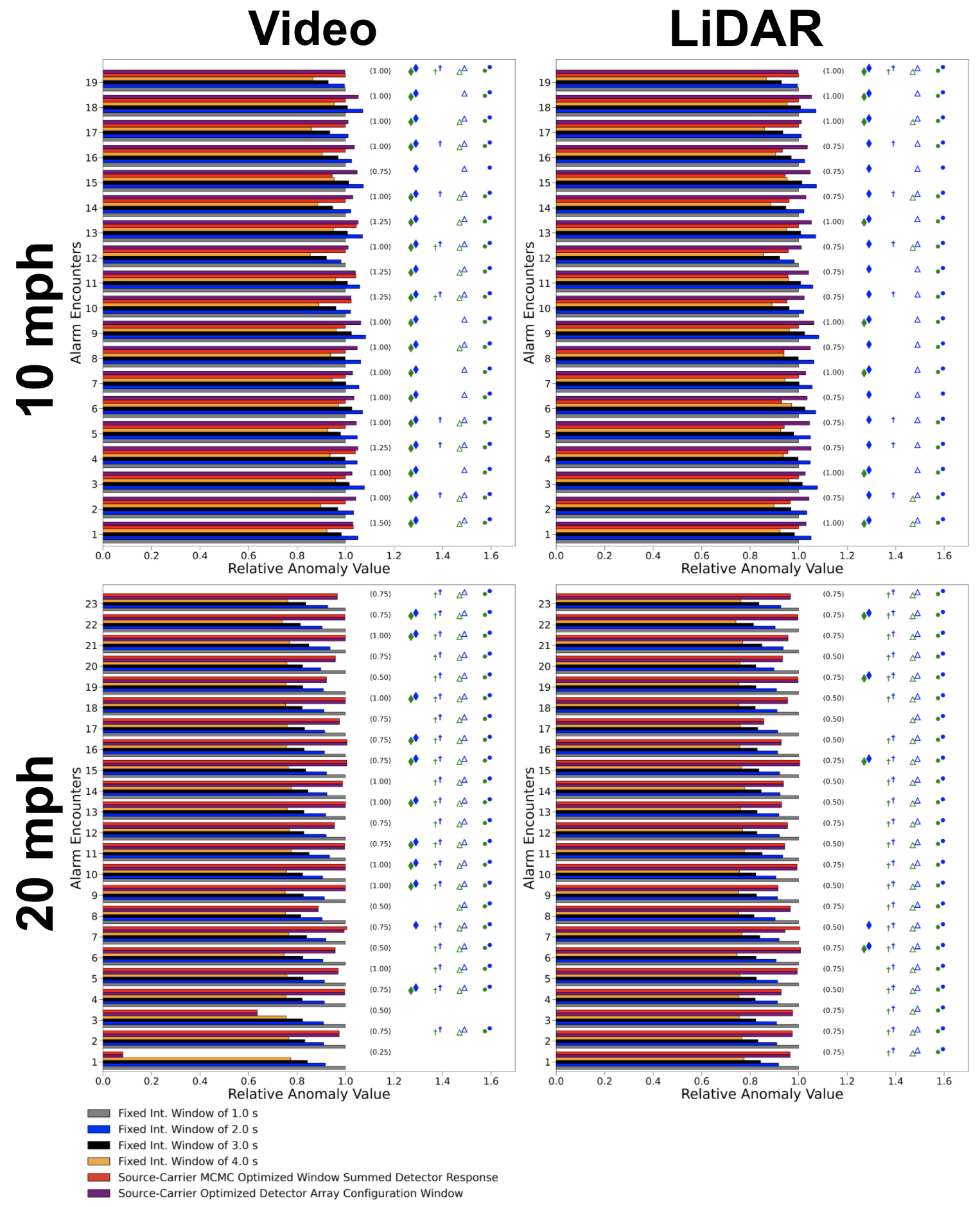}
    \caption{Maximum relative anomaly values for the 10~mph scenarios from \Fref{fig:10mph_complex}a (top-left) and \Fref{fig:10mph_complex}a (top-right) using either track-optimized time-windows or fixed integration windows for both video and LiDAR, respectively. 
    The maximum relative anomaly values for the 20~mph scenarios from \Fref{fig:20mph_complex}a and \Fref{fig:20mph_complex}b are seen in the bottom-left and bottom-right images for video and LiDAR, respectively.
    In all of the images, the purple bars (red bars) indicate track-optimized windows using the 6~NaI detectors independently (summing the response of the detectors).
    The time duration for the summed response optimal window is provided in the parentheses.
    The grey, blue, black, and orange bars indicate a fixed integration window of 1.0, 2.0, 3.0, or 4.0~seconds, respectively.
    The blue (green) diamond, dagger, triangle, or filled in circles indicate the encounters where
    the optimal integration window yielded a higher anomaly value than a 1, 2, 3, or 4~second integration window, respectively, using the 6~NaI detectors independently (summing the response of the detectors).}
    \label{fig:10mph_complex_opt}
\end{figure*}

The top-left and top-right images of \Fref{fig:10mph_complex_opt} show the results from all of the alarm encounters when LEMURS was traveling 10~mph for both video and LiDAR, respectively. 
For example, in \Fref{fig:10mph_complex_opt} (top-left), Alarm Encounter 16 shows the results of applying a spectroscopic analysis using either the track-informed optimal configuration of detectors or the track-informed optimal integration window calculated from using the summed response of the detector array compared to different fixed integration windows. 
In this alarm encounter, the optimal integration window using the 6~NaI detectors independently produced the highest relative anomaly value compared to both the optimal integration window found using the summed response of the 6~NaI detectors and the different fixed integration windows; whereas, the optimal integration window found using the summed response of the 6~NaI detectors has a larger relative anomaly value compared to the 1, 3, and 4~second fixed integration windows. 
Altogether, with video trajectories (\Fref{fig:10mph_complex_opt} (top-left)), using the 6~NaI detectors individually (summed response) and MCMC to produce the track-informed optimal integration window yielded an optimal window in 8 (2) of the 19 trials. 
For LiDAR trajectories (\Fref{fig:10mph_complex_opt} (top-right)), the results show that the track-informed integration window is the optimal window in 6 (1) of the 19 trials for optimal configuration of detectors (summed response).
These results from this scenario for video and LiDAR aligns with the hypothesis that for a given alarm encounter there exists an optimal configuration of detectors that will maximize SNR compared to summing the response of all the detectors together. 
In addition, the findings suggest the optimal configuration of detectors can improve detection sensitivity on a mobile system relative to fixed integration windows using video and LiDAR.

In the alarm encounters where the track-informed integration window did not produce the maximum anomaly value, it is thought that the anisotropic attenuation from both LEMURS and the source carrier is sufficient enough to cause the assumptions of the analysis to fail. 
For example, Alarm Encounter 17 in the top-left image of \Fref{fig:10mph_complex_opt}, which corresponds with the alarm encounter displayed in \Fref{fig:10mph_complex_single_encounter}a, shows that a fixed integration window of 2~seconds produces the maximum relative anomaly value.
In \Fref{fig:10mph_complex_single_encounter}a, the effect of anisotropic shielding from both LEMURS and the source carrier is seen in the count-rate data at the beginning and end of the radiological alarm.
In this case, both the optimal integration windows found using either the 6~NaI detectors independently or the summed response of the 6~NaI detectors produce integration windows that capture the period of time from about 16 to 17~seconds during the time of closest approach; however, both of these optimal windows fail to capture the period of time from about 17 to 17.5~seconds which still has an elevated count rate compared to the background rate due to anisotropic shielding (figure not shown).
A 2~second fixed integration window is better able to capture this effect, which leads to a larger relative anomaly value. 
With LiDAR-based trajectories, since orientation information is available for tracked objects, certain assumptions about the amount of anisotropic attenuation present from the objects were made to mitigate this effect.
However, the effects of anisotropic attenuation are not accounted for in the model generation of video trajectories, but video track-informed integration windows produced larger anomaly values in more alarm encounters compared to LiDAR-based trajectories. 
This is mainly due to the higher uncertainty with video-based trajectories compared to LiDAR-based trajectories, which enables a wider parameter space for MCMC to explore when determining the optimal integration window. 

These results imply that the anisotropic attenuation modeling does not capture all of the anisotropic attenuation present from tracked objects during the alarm encounter. 
Higher fidelity modeling of the vehicle along with the vehicle intrinsics has potential to improve the attenuation modeling. 
However, there is large uncertainty in the source location within the vehicle and the make and model of vehicles vary greatly. 
Both of these factors will affect the attenuation imposed by occluding material(s) in the vehicle as LEMURS and a source carrier pass by each other.
In addition to anisotropic attenuation modeling, since both the video and LIDAR track-informed integration windows did not produce the maximum anomaly value in a majority of the alarm encounters, these findings also suggest that the generated directional response matrix ($\epsilon$) does not capture all of the anisotropic attenuation from LEMURS itself, which could be improved with a more detailed model of LEMURS.
Overall, with a better understanding of how to account for different uncertainties associated with anisotropic attenuation from tracked objects as well as a more detailed LEMURS model, detection sensitivity should be improved, but a detailed characterization of both anisotropic attenuation from tracked objects and LEMURS is beyond the scope of this publication.

\setlength{\tabcolsep}{0.5pt}

\subsection{20 mph Scenario}
\label{sec:20mph_opt_win}
The spectroscopic analysis was also applied to the alarm encounters where both the source carrier and LEMURS were traveling at 20~mph relative to each other to further explore the optimal integration window analysis in more transient alarm encounters.

In the bottom-left and bottom-right images of \Fref{fig:10mph_complex_opt}, the results of applying this analysis to all 23 alarm encounters for video and LiDAR are shown, respectively.
In \Fref{fig:10mph_complex_opt} (bottom-right), the track-informed integration window produces the maximum anomaly value in 11 (10) out of 23 trials compared to different fixed integration windows using an optimal configuration of detectors (using the summed response of the detector array).
There was inconsistent tracking with alarm encounter 1 and 3 causing degraded detection performance.
The LiDAR-based track-informed integration window produces the largest anomaly value in 5 (4) out of 23 alarm encounters using an optimal configuration of detectors (using the summed response of the detector array) shown in \Fref{fig:10mph_complex_opt} (bottom-right). 
In this scenario, the time of closest approach between LEMURS and the source carrier is shorter and we are using an integration time of 0.25~seconds, which results in a 1/r$^2$ profile of the radiological data that is less distributed in time compared to the 10~mph case. 
This reduces the contribution of background when a spectroscopic analysis is performed.
As a result, the anomaly values using the summed response of the detector array are more comparable to the optimal configuration of detectors. 
However, the optimal configuration of detectors still produces a larger anomaly value compared to the summed response in 15 (18) of the 23 alarm encounters in \Fref{fig:10mph_complex_opt} (bottom-left) (\Fref{fig:10mph_complex_opt} (bottom-right)). 
Thus, the results in this faster scenario further highlight the advantage of identifying an optimal configuration of detectors for an alarm encounter over summing the detector array to improve detection sensitivity.  

Even though the track-informed integration windows did not always produce the maximum anomaly value, the results from this analysis demonstrate that our track-informed integration approach can better inform integration times by adapting to the dynamics of the scene and relative motion of objects in a scene for both video and LiDAR to improve detection sensitivity.
In the top-left and top-right images in \Fref{fig:10mph_complex_opt}, the maximum anomaly value in a majority of the alarm encounters was generated using a 2~second fixed integration window; whereas in the 20~mph scenario, a 2~second window only produced the maximum anomaly value in 1 alarm encounter.
In this scenario, a 1~second fixed integration window generated the maximum anomaly value in the majority of cases.
Multiple fixed integration windows can be run in tandem in an effort to maximize detection sensitivity.
However, it is not possible to account for the countless changes that can occur in a scene that could impact detection sensitivity
with a fixed integration window.
For example, if a source-carrying vehicle is stopped at a light near LEMURS for more than 10~seconds, a 1~ or 2~second fixed integration window will give lower anomaly values compared to a longer integration window.
This is the advantage of using the track-informed integration window which adapts to the scene and motion of objects.
\label{sec:results}
\section{Conclusion}
\label{sec:conclusion}
On mobile detection systems, alarm encounters are transient and localization has to be performed quickly and efficiently.
Source-object attribution enables new capabilities on a mobile platform by providing automatic associations between objects in the scene and the radiological data.

This work demonstrates that situational awareness can be improved in a mock urban environment on a mobile detection system in the presence of dynamic sources using SLAM and video or LiDAR-based trajectories. 
The findings show that video or LiDAR offer similar tracking performance which enable effective rejection of tracks that are inconsistent with the radiological data.
This performance is seen for various LEMURS and source carrier vehicle speeds. 
This work also explored performing the source-object attribution analysis using a navigational system that relies on GPS. 
The findings from this study demonstrate that if there is a strong GPS signal and the LEMURS vehicle speed is low ($\sim$10~mph) then using INS pose estimates to perform tracking and attribution offers similar capabilities to using SLAM with video and LiDAR-based trajectories. 
With faster LEMURS speeds ($\sim$20~mph) and INS, SLAM might be needed to generate more accurate pose information in situations where the distance information of the sensor is not directly known and is inferred.

Additionally, the findings from this work suggest that an optimal configuration of detectors does exist to improve the detection sensitivity of a detector array compared to either summing the response of the detector array or different fixed integration windows.
The track-informed integration windows from video and LiDAR trajectories are able to adapt to the dynamics of the scene and improve the anomaly value, proxy for detection sensitivity, in two different transient alarm encounter scenarios.
Overall, the conclusions from this analysis demonstrate that source-object attribution does improve situational awareness on a mobile platform system and suggests that detection sensitivity can be improved as well.

Future work is needed to fully explore using an INS in urban environments to produce pose estimates in a world-fixed frame and the effect this would have on tracking and attribution.
While the pose estimates generated using the INS did not impact attribution performance using LiDAR, it is thought in a true urban environment with buildings there will be potential for degraded tracking and attribution performance.
Urban environments are full of buildings, which will interfere with satellite signals more than the trees in the mock urban environment considered in this work. 
Additionally, since source-object attribution enables a new paradigm for source localization, future work will explore how source-object attribution might be considered in the design of detector arrays, or most effectively generalized to be robust to unknown detector configurations. 

Finally, future work will investigate scenarios with different source carriers, such as pedestrians, and scenarios where attenuating objects are present between the source object and detectors.
\label{sec:conclusions}
\section{Acknowledgements}
This work would like to acknowledge Adam Glick, Ivan Cho, Raymon Cheng, Kyle Bilton, Mark Bandstra, and Victor Negut for contributing to this work.\label{sec:acknowledgements}

\bibliographystyle{IEEEtran}
\bibliography{ref/references}

\end{document}